\documentclass{sigchi}

\usepackage{balance}  
\usepackage{graphics} 
\usepackage{times}    
\usepackage{url}      

\def\pprw{8.5in}
\def\pprh{11in}

\usepackage[pdftex]{hyperref}
\hypersetup{
colorlinks,
citecolor=black,
filecolor=black,
linkcolor=black,
urlcolor=black,
breaklinks=true,
}

\newcommand\tabhead[1]{\small\textbf{#1}}

\makeatletter
\def\@copyrightspace{\relax}
\makeatother

\begin{document}

\title{Matching or Crashing? Personality-based Team Formation in Crowdsourcing Environments}

\author{
  \alignauthor Ioanna Lykourentzou\\
    \affaddr{Luxembourg Institute of Science and Technology}\\
    \email{ioanna.lykourentzou@list.lu}\\
  \alignauthor Angeliki Antoniou\\
    \affaddr{University of Peloponnese}\\
    \email{angelant@uop.gr}\\
  \alignauthor Yannick Naudet\\
    \affaddr{Luxembourg Institute of Science and Technology}\\
    \email{yannick.naudet@list.lu}\\
}

\maketitle

\begin{abstract}
``Does placing workers together based on their personality give better performance results in cooperative
crowdsourcing settings, compared to non-personality
based crowd team formation?"
In this work we examine the impact of personality compatibility on the effectiveness of crowdsourced team work. Using a personality-based group dynamics approach,
 we examine two main types of personality
combinations (matching and crashing) on two main types
of tasks (collaborative and competitive).Our experimental results show that personality compatibility significantly affects the quality of the team's final outcome, the quality of interactions and the emotions experienced by the team members. The present study is the first to examine the effect of personality over team result in crowdsourcing settings, and it has practical implications for the better design of crowdsourced team work.\end{abstract}

\keywords{
	Crowsourcing; team formation; personality-based matching \newline
}

\section{Introduction}

Efficient team collaboration is a decisive factor for the success of any group project. Team formation, i.e. the selection of which individuals will become part of the team is one of the most critical steps in this process. Among the factors that play a role in a successful team formation are the individual team members' personalities. Indeed, as literature indicates, teams with matching personalities cooperate more efficiently compared to those teams where the participants' personalities do not match, or even crash \cite{Belbin:2010}. This knowledge, if properly exploited and applied at large-scale, could be valuable for enhancing team output in crowd work settings.

Cooperative crowdsourcing is a relatively new form of crowdsourcing, in which crowd workers interact to accomplish tasks either collaboratively or competitively, in contrast to typical crowdsourcing applications that comprise independent worker effort. A cooperating crowd can nonetheless accomplish more complex, interconnected tasks, due to the combination of various skills and knowledge backgrounds, with example applications including ideation contests, knowledge synthesis, collaborative problem solving and citizen science, to mention just a few {\cite{{Franzoni:2014,Kittur:2010}}. Yet, similarly to micro-task-based crowdsourcing, cooperative crowdsourcing also faces quality concerns and although certain works try to improve crowd team efficiency, most often through examining the proper incentives to give \cite{Hossain:2012}, very few works to-date exploit group dynamics and none exploits personality compatibility among the crowd team members.

Research on group dynamics and individual personality is vast in the fields of social psychology and personality psychology respectively. The formation of groups by matching the individual members' personalities is a field that has been studied less, mostly due to its multi-factorial nature (i.e. personality factors, situational factors, interaction factors) \cite{Furnham:1992}. Nevertheless, certain approaches and assessment tools can be found in this direction (see related literature) and this psychological knowledge can be exploited to assist group formation in cooperative crowdsourcing. These approaches need to be carefully examined prior to any application on crowd environments, due to the differences between crowd teams and the teams typically examined in social psychology. Indeed, whereas the typical team settings examined by group dynamics studies are mostly face-to-face, the people in cooperative crowdsourcing need to work from a distance and mostly asynchronously (due to the different time zones, availabilities and work schedules).
Therefore, the idea of bringing together crowd workers based on their personalities is something that needs to be tested, and this is exactly what this work is about. 

In this paper we examine the impact of personality compatibility on the effectiveness and final output quality of crowd teamwork.
Based on the DISC personality test \cite{Marston:1979} and the interactionist approach (both borrowed by group dynamics studies) we examine two main types of team personality compatibility, crashing and matching, on two cooperative task types, collaborative and competitive, applied on advertisement development. Our study has practical implications for the design of cooperative crowdsourcing and it can be used by task designers as a relatively simple (requiring an initial personality test) way of ensuring high-quality group results. 
\vspace{-0.5em}
\section{Related Work}
\label{related}
Crowdsourcing is a successful paradigm, with high commercial, educational and academic potential. Most commercial crowdsourcing applications are based on micro-tasks, which are given to independent workers and do not require cooperation \cite{Vaish:2014,Bernstein:2011}. Examples of this kind of crowd work include text translation, sentiment analysis, audio transcription and image recognition. Recent research explores using crowdsourcing for more complex tasks (e.g. \cite{Lasecki:2013}), which are often interdependent, of subjective nature and based on worker cooperation \cite{Kittur:2010}. Examples of such tasks include news article writing, product design or collaborative software development. 

The main concern that often hinders trust in crowdsourcing, either micro-task or cooperative-based, is the final outcome's low quality. A line of works explores the use of automated means to improve quality without exceeding the available task's budget. Indicatively, Karger et al. \cite{kargerBudget} use plurality optimization mechanisms for finding the optimal number of workers to allocate per micro-task, in order to ensure high quality while minimizing task cost. Other works apply preprocessing to filter out low-quality workers, based on reputation mechanisms, screening mechanisms \cite{Downs:2010}, pre-qualification tests, or golden data \cite{Josang:2007}. Post-processing is also applied to refine and evaluate task quality after the tasks are completed \cite{whitehill:2009}, or while they are being processed \cite{Ramesh:2012}. 

Another line of studies point out that enhancing crowd work quality needs a change of viewpoint: from considering workers as homogeneous, interchangeable units (typical crowdsourcing model) to taking into account the human factor, i.e.the emotional and cognitive personal characteristics of the workers.
Motivation is the factor most extensively examined and many works have found significant correlations between various incentives and task output quality especially as far as creative or innovative tasks are concerned (indicatively \cite{Hossain:2012}). Morris et al. \cite{Morris:2013} use priming to increase the performance output of workers in creative crowdsourcing tasks. Their results confirm that this technique helps improve worker performance. Sampath et al. \cite{Sampath:2014} use cognitive-inspired features in task design as a technique to improve the quality of the crowdwork. 

Few works also explore the use of personality in crowdsourcing. Indicatively, Kazai et al. \cite{Kazai:2011,Kazai:2012} examine the quality of the workers' output in relation to their personality traits.
Their results confirm a strong correlation between worker personality traits and their work-related traits (tasks completed, task completion time and accuracy). Other works in this direction also use personality aspects to predict differences between worker stereotypes (competent/incompetent, meticulous/sloppy etc.)\cite{Eickhoff:2011,Vuurens:2011}.
The above works are in line with the present study regarding the importance of taking into account the personality of crowd workers for their better selection and allocation to the tasks. However, most current studies focus on individual workers, while the present work focuses on the use of personality for the composition of worker teams, therefore targeting not only worker-to-task but also worker-to-worker matching.

From a psychology perspective, literature either focuses on the individual or on the group. Regarding the individual, the individual characteristics both cognitive and biological can be already effectively measured (e.g. conservatism – \cite{Wilson:1973,Brewin:1988,Eysenck:1967} sensation-seeking – \cite{Zuckerman:1979} etc.). Also, valid theories and tests that categorize people's personality traits as individuals, like Holland's 6 personality types \cite{Holland:1973}, Costa and McCrae's \cite{Costa:1992} NEO-PI-R five factor analysis, Cattell's 16 factors \cite{Cattell:1977}, Myers-Briggs Type Indicator \cite{Myers:1985}, or Eysenck's supertraits \cite{Eysenck:1975} study people as individuals and not as part of groups. 

Regarding groups, there is extensive literature in personality and social psychology about groups and member's behavior. Less research however exists on which specific personalities one can bring to a group in order to increase efficiency and how a specific individual with certain personality traits will behave once in a group. Relevant to this, the person-situation debate in psychology (whether a person's personality or the situation is the main determinant of her behavior and performance) can be summarized by three main theories \cite{Furnham:1999}: \textit{Trait theories} support that personality is the main factor; \textit{Situational theories} support that the situation is the main behavior factor; and \textit{Interaction theories} support that behavior is a synthesis of the two \cite{Carver:1996}.
In this work, we decided to follow the interactionist approach as it includes more factors, it is supported by long-term research data (e.g. see the 15-year long review by \cite{Pervin:1984}) and its usefulness has been already demonstrated in HCI \cite{Nov:2013}. Consequently, we rely on variants of both situational elements (e.g. nature of the task) and individual personality traits in order to extract the variables affecting group performance that will be used in this paper. 

Regarding personality factors, there are few theories and tools that study the individual as a part of the group. The DISC personality test \cite{Marston:1979} identifies four main types of group members: \textit{2 leader types} with high \textbf{D}ominance (task-oriented, focus on task completion) or high \textbf{I}nducement (socio-emotionally oriented, focus on interpersonal relations) and \textit{2 non-leader types} with high \textbf{S}ubmission (socio-emotionally oriented) or high \textbf{C}ompliance (task-oriented). According to Belbin's approach \cite{Belbin:2010} effective teams include people of 8 different types (Chairman, Shaper, Plant, Monitor evaluator, Resource investigator, Teamworker, Company worker, Completer) and a team is successful if all of the above roles are covered. 

Regarding situational factors, the literature emphasizes the importance of the task's nature, i.e. that a person's work behavior is highly related to the task that the person is involved in. After reviewing the Steiner \cite{Steiner:1972} task typology and other available literature \cite{Lord:1997}, we identified 7 main task types. \textit{Task type I} denotes whether a task can be divided to further subtasks (values: 1. Divisible-existence of subtasks, 2. Unitary-no subtasks). \textit{Task type II} denotes whether the team focuses on the quality or the quantity of the task (values: 1. Maximizing-importance placed on quantity, 2. Optimizing-importance placed on quality). \textit{Task type III} denotes the mechanism used by the team to combine the contributions of its individual members (values: 1. Additive-individual inputs are added, 2. Compensatory-group product is the average of individual judgments, 3. Disjunctive-product is selected from pool of individual judgments, 4. Conjunctive-product is a synthesis of all member contributions, 5. Discretionary-group can decide how individual inputs relate to group product). \textit{Task type IV} denotes the way of cooperation among the team members (values: 1. Collaborative-commonality of interests, 2. Competitive-conflict of interests, 3. mixed motive-both common and conflicting interests). \textit{Task type V} denotes the level of difficulty of the task (values: 1. Easy, 2. Difficult). Task type VI denotes the task's duration (values: 1. Short, 2. Long). \textit{Task type VII} denotes the subjectivity of the task (values: 1. Intellective-a correct answer exists, 2. Judgmental-no demonstrably correct answer). 

In addition to the 7 task types, another very important factor of group productivity is the \textit{number of group members}. Due to the Ringelmann effect \cite{Kravitz:1986}, there is an inverse relationship between the number of people in the group and the individual performance. Possible values are: 2-7, 8-9 or 9-16 members, since as the literature shows, significant qualitative differences are observed in the behavior of groups below, around and over 8 people \cite{Steiner:1976}. Also, the \textit{amount of control given to the group leader} in a given situation seems to play a crucial role. According to contingency theories, the leader's type and the situational control she might have, affect the group's outcome \cite{Fiedler:1983}. 
Last, groups seem to interact according to four main \textit{group interaction types} \cite{Furnham:1999}: 1. Interacting (natural processes occurring during face-to-face interactions), 2. Brainstorming (synchronous technique that encourages all ideas while withholding any criticism), 3. Nominal (both synchronous and asynchronous technique: members first work independently, then meet to discuss their ideas) and 4. Delphi (asynchronous technique, similar to nominal groups: members never meet, instead they first work independently, then they see other member ideas and work again alone).  
\section{Methodology}
\label{methodology}
\subsection{Research Design}

From the above, we identify 11 main personality and situational elements that affect group performance (Table~\ref{tab:table1}). Given their potential values, these elements give rise to a significant number of possible experimental combinations ($>23000$). In this section, we describe the decisions taken, in regards to the values of these elements, which lead us to our specific research design, and suited the context of personality-based matching in cooperative crowdsourcing.   

The first decision pertains to the personality assessment tool that will allow the extraction of the individual personalities of the crowd workers (Individual Personalities element). From all the available theories and tests, the ones that have direct relation to team work and not simply individual assessments were chosen. In particular, the DISC test focuses on the way that different group members will interact with each other and the roles that they will play inside the group. In addition, the DISC test differentiates between 4 main types (with fluctuations in the proportion of the different dimensions), whereas Belbin (the other candidate test) identifies 8 main types present in ideal teams. Thus, it was decided to use the DISC test in the current study mainly for practicality reasons, since it leads to smaller and easier to handle worker groups. In a future work however, the Belbin test will be also used. 
Based on the DISC test, two types of groups were formed, in regards to personality compatibility: 
\begin{itemize}
\item Groups of \textit{matching personalities}. They consisted of one \textbf{D}ominant personality, one \textbf{I}nducement personality, one or two \textbf{S}ubmission personalities and one or two \textbf{C}ompliance personalities. This group type included all the DISC types while avoiding the presence of two similar types of leaders.
\vspace{-1.2em}
\item Groups of \textit{crashing personalities}. They consisted of either more than one leaders of similar type (usually \textbf{D} types).
\end{itemize}

Following the interactionist approach, apart from personality elements, we also incorporated situational elements, i.e. specific task types (\textit{elements: task type I-VII}). For the context of this research, it was decided to examine group performance under competition as well as collaboration. Thus, each of the basic groups was further divided into two more categories(\textit{task type IV}: Collaborative/Competitive, \textit{task type III}: Conjunctive/Disjunctive):
\begin{itemize}
\item \textit{Collaborative}, where workers co-create a concept.
\vspace{-0.7em}
\item \textit{Competitive}, where workers compete for the best concept.
\end{itemize}
The remaining task type elements were kept stable across all worker groups. Specifically, since our particular research design aimed at crowdsourcing contexts, the task that the workers would accomplish needed to be of short duration to increase chances of task completion by the participants (\textit{task type VI: Short}) and fully computerized, so that people would be able to perform it without leaving their PC, from a distance and without time zone constraints. Also, since we aimed at a broad crowdsourcing worker pool the task should not require prior expertise (\textit{task type V: Easy}). Since we needed to examine the influence of personality, the task should not be routine or repetitive but rather creative, to allow the expression of the workers' personality. Being a creative task, there is no correct or incorrect answer (\textit{task type VII: Judgmental}). A judgmental task would further allow diverse group processes to emerge. In order not to interrupt the group creativity processes, the task was also chosen to be unitary (no subtasks) (\textit{task type I: Unitary}). Since we were interested in the quality of the final group outcome, the task's objective should focus on quality rather than on quantity (\textit{task type II: Optimizing}). 

The present study operated with groups of 5 people (\textit{\# Team members: 5}), due to the assessment tool used (minimum 4 members) and in line with research findings showing that an effective group should not exceed 8 members \cite{Steiner:1976}.  

In regards to leadership we did not impose any type of leadership and each group was allowed to perform as its members wish (\textit{Leader control: Low}). Despite the fact leaders were identified from the initial personality test, group members were allowed to interact freely with each other and without knowing who the leader is, in order to see the actual group dynamics and not the ones we had predicted before the interaction. This choice was precisely meant to allow us observe whether leadership would emerge and under which personality combinations.

The chosen group interaction type was Delphi, since it is a type of nominal group approach and nominal groups seem to provide better results than other types (interaction and brainstorming) \cite{Taylor:1958}. The asynchronous nature of Delphi would also allow the interaction of participants from different time zones and work schedules, in a sequential rather than simultaneous manner, which is found to be better for uncertain, subjective tasks, like the ones used in this research \cite{Andre:2014}. The research design presented above, resulted in the following 4 experimental conditions:

\begin{enumerate}
\item \textbf{CR/CM: Crashing Competitive.} A group with crashing personalities, working on the task competitively. 
\item \textbf{CR/CL: Crashing Collaborative.} A group with crashing personalities, working on the task collaboratively. 
\item \textbf{M/CM: Matching Competitive.} A group with matching personalities, working on the task competitively.
\item \textbf{M/CL: Matching Collaborative.} A group with matching personalities, working on the task collaboratively.   
\end{enumerate}
\subsection{Research Hypotheses}
Given our basic question: ``Does team formation based on personality matching give better performance results in cooperative crowdsourcing settings, compared to non-personality based matching?", our two main research null hypotheses are: 

\begin{enumerate}
\item \textit{$H_{o1}$. Quality of final outcome}. The quality of the final outcome of the group work will not have significant differences among the 4 experimental conditions. Especially the matching personality conditions are not expected to outperform the crashing personality conditions. 
\item \textit{$H_{o2}$. Group effectiveness and emotions}. The quality of the perceived group effectiveness and emotions will not have significant differences among the 4 experimental conditions. Especially the participants of the matching personality conditions are not expected to work more efficiently and experience less negative emotions (motivation, satisfaction, frustration, confidence, etc.), compared to the participants of the crashing personality conditions. 
\end{enumerate}

The above represent the two fundamental, generic hypotheses that this research dealt with. Additional sub-hypotheses have been identified and dealt with, which are not discussed in this section for reasons of space and readability. Part of these is presented in the results section, and the rest as part of the future work in the discussion section. The identified hypotheses will be analyzed qualitatively and quantitatively.
\begin{table}[ht!]
  \centering
  \begin{tabular}{|p{2.5cm}|p{5cm}|}
    \hline
    \tabhead{Variable} &
    \multicolumn{1}{|p{0.3\columnwidth}|}{\centering\tabhead{Value}}\\
    \hline
    Individual Personalities & \textbf{DISC-based (4 types)} \\
     	& Belbin-based (8 types) \\ 
    \hline
    \# team members & \textbf{2-7} \\
    	& 8-9 \\
    	& 9-16 \\ 
    \hline
    Task type I & Divisible \\
    			& \textbf{Unitary} \\
    \hline
	Task type II	& Maximizing \\
					& \textbf{Optimizing} \\
	\hline    
	Task type III	& Additive \\
					& Compensatory \\
					& \textbf{Disjunctive} \\
					& \textbf{Conjunctive} \\
					& Discretionary \\
	\hline    
	Task type IV	& \textbf{Collaborative} \\
					& \textbf{Competitive} \\
					& Mixed motive \\
	\hline    
	Task type V		& \textbf{Easy} \\
					& Difficult \\
	\hline    
	Task type VI	& \textbf{Short} \\
					& Long \\
	\hline    
	Task type VII	& Intellectual \\
					& \textbf{Judgmental} \\
	\hline    
	Group interaction	& Interaction \\
						& 	Brainstorming \\
						& Nominal \\
						& \textbf{Delphi} \\
	\hline    
	Leader control		& \textbf{Low} \\
						& High \\
	\hline   
	Other environmental variables & Known variables to affect group work (i.e. affecting productivity, cohesiveness, creativity, etc.) will remain stable during the different experiments. The same guidelines for group work will be followed through the study. \\
\hline   
  \end{tabular}
  \caption{Variables affecting group performance (values used in present research in bold)}
\vspace{-1em}
  \label{tab:table1}
\end{table}
\vspace{-0.3em}
\subsection{Experiment Implementation}
\subsubsection{Task Description}
According to the requirements identified in our research design, we decided to use the task of \textit{cooperative advertisement creation}. Specifically, as also shown by Dow et al. \cite{Dow:2011:PDS:1978942.1979359} an advertisement task fulfills certain of our key criteria like: short duration, no expert and previous knowledge requirement, ability to express creativity and ability for both objective and subjective measurements of quality. According to this task, groups with matching or crashing personality combinations would be asked to create the advertisement campaign of a new product, either competitively or collaboratively. Generally speaking, a product's campaign can consist of many elements, like slogan, scenario, music, logo, etc. It can also vary depending on the broadcasting medium (television, radio, Internet etc.). To keep the task short, here we choose to ask workers for the product's slogan (text up to 50 words) and scenario (text up to 150 words) aimed for TV broadcasting. The product to advertise was a new fictive coffee beverage, called ``sCOPA". Coffee was used, after reflecting among various candidate products, because it is a product likely to be known to people across the globe, with rather neutral belief connotations (e.g. religious, political, etc.), and without being exclusively associated with any particular brand (as it would be the case e.g. for specific soft drink products). 
The task was implemented in two versions:
\begin{itemize}
\vspace{-0.4em}
\item \textit{Competitive task version.} The group is asked to create the final advertisement by selecting one single campaign, among the ones proposed by its individual members. 
\vspace{-0.4em}
\item \textit{Collaborative task version.} The group is asked to create the final advertisement by combining the campaign ideas proposed by its individual members. The group members are free to take ideas one from the other, and change the original texts. 
\vspace{-0.3em}
\end{itemize}  
\subsubsection{Crowdsourcing workflow}
We used the crowdsourcing platform CrowdFlower.com mainly for its breadth of worker sample (access to 5M workers from 154 countries in over 50 labor channels). Ethics approval was obtained and all legal requirements for data protection were fully followed. Participants were informed about the academic nature of the experiments and their legal rights. Following this, the implementation of the experimental design was conducted in 3 rounds.
\paragraph{Round 1. DISC personality test}
The 1st round was an open crowdsourcing task, where workers were invited to take the DISC personality test. This task paid 1\$. 295 workers from 59 different countries participated in this round. Each worker was asked if she would like to participate in the next rounds (subject to selection based on her profile) and, in case of a positive answer, to provide us with a contact email. 
\paragraph{Round 2. Individual advertisements}
In the 2nd round the workers who stated interest to participate were invited to make an individual advertisement (slogan and scenario as described above) about the sCOPA coffee product, through a dedicated CrowdFlower job that paid 1\$. They were instructed that their \textit{``ads should be original with a clear market value, using simple, understandable and honest messages and emphasizing on the unique aspects of the product"}. These instructions were meant to align worker contributions with the final outcome quality axes that we intended to measure at the end of the experiment (see Evaluation Metrics Design sub-section). 185 workers participated in this round.
\paragraph{Round 3. Cooperative advertisement creation}
The 3rd and most important round of the experiment consisted of selecting the workers and placing them into the groups. Four distinct types of worker groups were created, according to our 4 experimental conditions. Selected workers were invited by email. Each group comprised 5 workers, who were given a link to a Google document, on which they would work to create the final sCOPA advertisement. This document contained 3 parts: 1) Task instructions, 2) the 5 individual advertisements created by the individual team members in Round 2, and 3) document space to host the final group advertisement. 
The competitive groups were instructed to read the individual advertisements, discuss and select the best one, without any changes in the slogan or scenario. The collaborative groups were instructed to read the individual worker advertisements, discuss and create one new advertisement by merging, modifying and taking ideas from any individual advertisement they wanted. Workers of all groups were instructed to actively discuss and interact with the other people in their groups, for the final group outcome. The interaction was asynchronous, through threads of comments that the workers would add to the Google document. Each group had a working period of 5 days, to keep the task short. One day before the deadline each worker group was sent a reminder, inviting people to participate if they had not done so. 
To motivate participation, workers were paid based on their level of interaction with their groups (0.5-2\$), while an extra bonus was given to those groups that managed to make the final advertisement (1\$). 145 people participated to the 3rd round, split into 29 groups. 
\subsection{Evaluation metrics design}
Following our two hypotheses, we evaluated the: i) Final group outcome and the ii) Group effectiveness and emotions. This was a multidimensional evaluation process, where both quantitative and qualitative metrics were used. 
\paragraph{Final group outcome evaluation}
According to Hoffman \cite{Hoffman:2012} the successful ad: is creative, dramatizes and communicates the reasons to buy the product, is honest, is simple (one message is better than two), rhymes things, is possible, and looks for the product's Unique Selling Point. Based on this study, as well as on similar recent research developments on information and content quality \cite{Chai:2009,Knight:2005}, we defined five axes of final group outcome quality: \emph{1) Originality} (How original and creative is the advertisement?), \emph{2) Market Value} (How likely is it that the advertisement will attract customers?), \emph{3) Simplicity} (How simple and understandable is the message of the advertisement?), \emph{4) Honesty} (How honest is the advertisement?) and \emph{5) Unique Selling Point} (How well does the advertisement highlight the differences between this product and other similar products?). 

Although other dimensions could also be evaluated (\cite{Knight:2005, Chai:2009}), it was decided to keep the evaluation process simple and short, to facilitate the evaluators. The resulting questionnaire was given to an expert evaluator (advertisement industry professional) as well as to the 1250 crowd workers (50 workers per final advertisement), to also get the average user's opinion and capture the ``Wisdom of Crowds" effect (crowds can outperform the estimations of individual experts \cite{Forlines:2014}). Each worker assessed up to 5 advertisements to avoid working memory cognitive overload \cite{Miller:1956}. 
\paragraph{Group interaction effectiveness and emotions}
Following the 2nd hypothesis, participants after the 3rd round were given a questionnaire developed based on the emotions classification study by Pekrun's and colleagues' \cite{Pekrun:2011} and their Achievement Emotions Questionnaire.
It assessed the following: \emph{1) Motivation} (How motivated did the worker feel to participate to the group advertisement creation task), \emph{2) Stress} (How stressed or frustrated the worker felt during her interaction with the group), \emph{3) End result} (How satisfied the worker was with her group's end result), \emph{4) Communication quality} (How happy she was with the quality of communication among the group members), \emph{5) Sharing confidence} (How confident the worker felt to share her opinion with the group), \emph{6) Acceptance} (How well did the group welcome the worker's contribution), \emph{7) Opinion on cooperative tasks} and \emph{8) Interest for re-invitation} to similar tasks in the future. The questionnaire also included open-ended questions over the workers': \emph{9) Face-to-face behavior} (How the worker's behavior would be different in case the task was face-to-face), \emph{10) Process suggestions} (What would the worker change in the overall process) and \emph{11) Other comments}.
\section{Results}
\label{results}
\subsection{Overall sample statistics}
Overall, in a population of 295 workers that took the personality test of the 1st round, we observe the following:
\begin{itemize}
\item Leader types: \textbf{D}: 42.71\%, \textbf{I}: 9.15\%, \textbf{D/I}: 4.41\%
\item Non-leader types: \textbf{S}: 13.56\%, \textbf{C}: 13.90\%, \textbf{S/C}: 4.07\%
\item Mixed (all other combinations): 12.20\%
\end{itemize}

As it can be observed, the crowd worker population is not normally distributed, but there is a higher percentage of Leader types (56\%) versus non-leader types (31,46\%). Thus the probability of having a randomly selected team with more than one leader is high. This observation strengthens the significance of our results, since in case our hypotheses are verified, this would mean than a team formation which does not take into account personality compatibility risks a sub-optimal result. Also, this made the selection process more challenging since we needed to balance matching and crashing group populations. From the 185 people who participated in Round 2, 145 were invited to the 3rd round (in order to have a balanced number of teams). In the end, team formation was as follows: 29 groups, 5 workers each, of which 6 were CR/CM, 7 CR/CL, 7 M/CL and 7 M/CM.
\subsection{Observations on Group behavior}
To answer the question \emph{``did the groups behave as expected?"}, a qualitative analysis of comment logs was performed. Looking deeper into the group processes, a few very interesting patterns were revealed. Most groups did behave as expected. In fact, most tension among group members was built in the Competitive Crashing groups. Ironic comments were observed as people shouting (using capital letters): ``\emph{Dear [participant number removed] Thank you for your comment. I was waiting for such a remark. However, this could be the reason why somebody will NEVER FORGET this advertisement and for this reason REMEMBER TO BUY sCOPA COFFEE THE NEXT TIME HE VISITS A SUPERMARKET OR A COFFEE SHOP}".
Apart from the sharp comments, the competitive crashing groups were also spending a lot of time discussing the processes to follow, without easily reaching an end result. In one case, the group did not reach a final decision at all. Participant comments were revealing: \emph{``Apart from that, it seems that we are unable to agree on one advert being a perfect winner"} \dots \emph{``I'm beginning to feel like this is really a psychology experiment, and they want to see what we will do about people not participating"} \dots \emph{``Oh well, that didn't go as planned, did it? I think there will be a LOT for the academics to draw out from this experience!"}. 

On the other side, we observed the efficiency of the matching groups and especially the Matching Collaborative groups, which not only seemed to easily reach a group decision, but also created a positive and encouraging atmosphere: \emph{``To point it out again: Good job, team!"} \dots \emph{``I hope everything is okay with what we've done, great job everyone and good luck!"} \dots \emph{``It was great working with you all"} \dots \emph{``Yes, good job team! Hope to work with you in the future.. =)"}. The participant comments are presented exactly as written by the participants, or with clearly indicated grammar, spelling or other corrections inside brackets. 

Further interesting observations can be drawn regarding leader behavior. First, in the absence of leaders it was difficult for the group to reach a decision and in fact in 2 out of 3 cases, the group did not reach a decision at all. 
Second, in those Crashing conditions where the D leaders did not have an active participation the groups tended to convert to Matching and work without tension. Occasionally, in the cases when the D leaders did not dynamically participate, other members came in charge. These people were either type I leaders (if this personality type was actively participating) or even non-leader type. Third, in cases of Crashing groups where only up to 3 people participated, even if they were all D leaders, they seemed to communicate efficiently. In these cases, it seems that the group size is crucial, meaning that strong leaders can cooperate as long as they are not more than 3 people in a group. Third when the leaders (D and I) participated, the group functioned well. In their absence, however, the group crashed, until the leaders took over again and the group regained control. Finally, in a group with a strong D leader and other mixed types, the mixed types adopted a non-leader approach and let the strong leader lead the group.  
\subsection{Observations on Worker behavior}
Our next set of observations seeks to answer the question: \emph{``Did the group members behave as expected?"}. Analyzing individual behavior in a group is a very challenging task, since human behavior does not follow predetermined paths. However, we tried to observe leader and non-leader behavior in the different groups and it seems that the majority of individuals behaved more or less as expected. Leader types seemed to lead the groups and non-leader types seemed to follow. We also observed differences between socio-emotional leaders (personality type I) and task leaders (personality type D). There were only 3 cases of groups with members of unexpected behavior, meaning that the people categorized as leader behaved as non-leaders and vice versa.  

The D personalities were dominating the group processes in most groups. Indicatively, D leaders with clear task orientation, determined the decision-making processes: \emph{``I put already all the slogan below, in the decision page… [it] will [be] easier for us"} \dots \emph{``Let's vote here. Reply with your vote (don't forget own ID). Last one to vote, do us a favour by copy-pasting the winner's (ID, Slogan, Scenario)"} \dots \emph{``Hello friends, I leave this comment to remind those who have not yet participated that have until September 7 to give their [vote, so] that all participate."}. 

Socio-emotional leaders (type I) were focusing on the group interactions, encouraging other members: \emph{``I will not comment mine, but I had a funny time doing it. congrats to all"} \dots \emph{``That's a great scenario [participant ID removed]. I like that it emphasises the concept that the coffee can be drunk either hot or cold. I took the liberty of adding a slogan to the scenario! Feel free to change it if you disagree!"} \dots \emph{``well i like the first idea since I was the one who wrote it, to be fair its not that good and it can use some adjustments, tell me what you think, and of all the ideas here i think No 5 is fairly good"}.

\subsection{Hypothesis 1 Evaluation (Quality of final outcome)}
\paragraph{Expert evaluation.} The groups' final advertisements were evaluated by an expert advertisement professional for their quality in regards to originality, market value, understandability, honesty and unique selling point (evaluation dimensions as explained above, Hoffman, 2012). The overall score of each advertisement (measured in a scale [0-50], i.e. the sum of scores of the 5 individual axes) was calculated and the score of the 4 experimental conditions were compared using a Kruskal-Wallis analysis. The expert's ratings reveal a superiority of the Matching Collaborative groups' advertisements (mean score 24.71 out of 50). The worst end results came from the Crashing Competitive groups (mean score 14 out of 50), followed by Crashing Collaborative groups (mean 14.28) and Matching Competitive groups (mean 17.43). The same pattern is also observed for each of the 5 individual quality axes.
The overall evaluation rating of the expert is depicted in Figure~\ref{Fig_hyp_1}. 

\paragraph{Crowd evaluation.} The crowd ($N=1250$) agreed with the expert regarding the higher quality outcome of the matching groups (especially in the collaborative task), compared to the crashing ones. This is statistically confirmed, with one-way ANOVA analyses for all quality axes (indicatively for the overall quality rating axis: $F(3,1397)=28.05$, $p<.001$ and similar results with highly significant $p$ values $<.001$ for the individual axes). We note nevertheless that the crowd consistently provided higher marks than the expert. Figure~\ref{Fig_hyp_1}, illustrates the average crowd ratings, next to the respective expert ratings. From the above, null hypothesis $H_{o1}$ is rejected. 

\begin{figure}[!h]
\centering
\fbox{
\includegraphics[width=0.9\columnwidth]{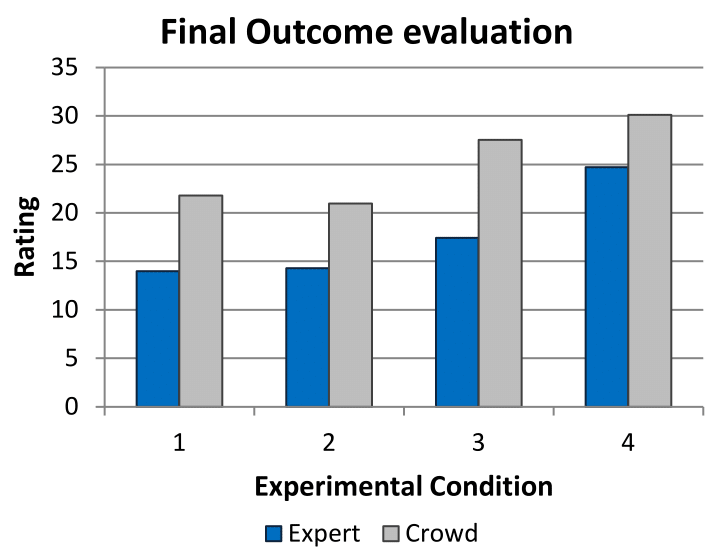}
}
\caption{Hypothesis 1 – Final Outcome Evaluation.}
\label{Fig_hyp_1}
\end{figure}

\subsection{Hypothesis 2 Evaluation (Group efficiency and emotions)}
Participant questionnaire answers were analyzed regarding hypothesis 2. Three statistically significant results were found, and their averages are depicted in Figure~\ref{fig_hyp_2}.

\paragraph{Communication Quality.} The participants of the Matching Collaborative groups reported the highest levels of satisfaction, followed by the Matching Competitive, Crashing Collaborative and Crashing Competitive groups. This result is statistically significant with $H=8.57$, $3$, $p<.05$ and fully in line with hypothesis 2. This quantitative result was further validated by qualitative analysis of participant comments. Indicatively, comparing the comments of a participant from a Crashing Collaborative group (\emph{``Unfortunately, i didn't have an active group in which a discussion could be properly held. Either they were a bit inactive, or their communication skills were a bit rusty. I tried leading the group, since they were all pretty pleasing with each others. Not many comments or edits were done. They would mostly throw their original idea, and that was about it."}), with the comment of a participant in a Matching Collaborative group (\emph{``great task! feels like i was working with team! i ll be very happy if i could do more tasks like this in future"}), we observe an obvious difference in opinion. This was also reflected in the word clouds (created using the Semantria software\footnote{Semantria, Lexalytics. https://semantria.com/}) to visualize the content of intra-group communication. Figure \ref{fig:figure3} provides one indicative word cloud example per group category. Similar patterns were found for most groups.

\paragraph{Stress Levels.} Individuals in the matching conditions and especially in the Matching Collaborative groups felt more relaxed and reported significantly lower stress levels than the crashing groups and especially the Crashing Competitive ones ($H=7.87$, $3$, $p<.05$). This result shows that in regards to stress levels, the personality compatibility (crashing or matching) does play a role especially when the task is of competitive nature. The same pattern is revealed through the qualitative analysis. Indicatively, although participants from the matching conditions do not report any stress issues and they are rather pleased with the overall experience, a participant from a Competitive Crashing group said: \emph{``\dots I was a little nerves [nervous] in case I was ``intruding" on regulars, but hopefully next time I'll have more confidence."}

\paragraph{End Result.} People in the matching groups and especially collaborative matching ones were more pleased ($H=8.23$, $3$, $p<.05$) with the group's final result than the people in the crashing conditions. This finding is also reflected in the qualitative data. However, it is interesting to see the in-between views of participants of the Competitive Matching conditions. Although they liked the other group members and enjoyed their interaction, the competitive nature of the task, left these participants with mixed feelings. Indicatively, a worker mentioned: \emph{``Team result is good but I'm [a] little disappoint[ed] because my hard work did not succeed. Overall I am happy that finally we have a deserving winner."} 

\begin{figure*}[ht!]
\centering
  \raisebox{-0.5\height}{\includegraphics[width=0.25\textwidth]{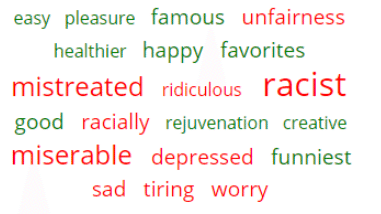}}%
\raisebox{-0.5\height}{\includegraphics[width=0.25\textwidth]{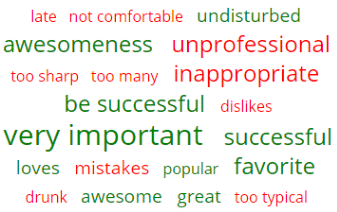}}%
\raisebox{-0.5\height}{\includegraphics[width=0.25\textwidth]{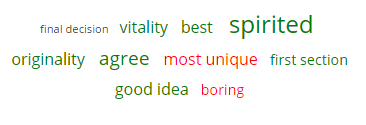}}%
\raisebox{-0.5\height}{\includegraphics[width=0.25\textwidth]{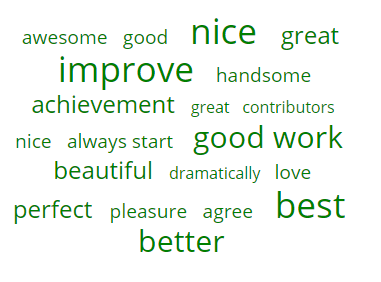}}%

\caption{Word clouds of team discussions. From left to right: Crashing Competitive, Crashing Collaborative, Matching Competitive and Matching Collaborative. Word/concept frequency indicated by word size. Positive sentiments in green, negative sentiments in red, neutral in black.}
\label{fig:figure3}
\end{figure*}

Finally, no statistically significant differences were found in regards to motivation, sharing confidence and acceptance. All workers reported that they were highly motivated to participate (mean=2.89, SD=0.18), confident to share their opinion (mean=2.81, SD=0.24) and felt relatively accepted by their group (mean=2.58, SD=0.41). All participants, with no statistical significance across the groups, expressed satisfaction with cooperative crowdsourcing tasks (mean=2.84, SD=0.21) and interest to be re-invited to similar tasks in the future. Finally, participants reported motivated to participate: \emph{``I enjoyed this task a lot, specially the third round. Although there were differences in opinions and moments when the member couldn't even agree to disagree, it was a fun and motivating experience."} They were also pleased with the fact that they could share their ideas with others: \emph{``it was a fun way of making and sharing opinions with others with the benefits of doing something important"}. 

Overall, most participants were happy with a cooperative crowdsourcing task: \emph{``a very creative way to make people working together"} \dots \emph{``I really liked this job and looking forward to participate further in such jobs!"} \dots \emph{``I'm very happy that I had the chance to participate in this test. I believe that it would be very interesting to see more task[s] like this in the future, tasks that make workers think and express themselves on different ideas. Thanks so much,"} \dots \emph{``Thanks very much for this task, it was the one I have enjoyed the most since I started doing tasks. Being able to be creative while having a chance to work with others was a really, really great experience."}

Although many people suggested that a synchronous communication would be beneficial for this kind of task and many believed that they would be more active in a face-to-face task, however, a few participants 
raised concerns: \emph{``\dots I would be less comfortable in a face to face situation"} \dots \emph{``I would be more quiet. I'm more confident when I'm writing. I was the first to comment on the Google document, but if this was a face to face task I believe I would listen to other people's opinions before speaking"} \dots \emph{``face2face would be easier (less pressure on language writing skills), but it would be less convincing, face2face required specific times and that can be a big problem"} \dots \emph{``if it was face-to-face task i would be more emotional because there wouldn't be time to calm down, if i don't like something"}. Thus, it seems that a synchronous or a face-to-face interaction would be better for some and not all participants. However, it is definitely worth exploring further in a future study.

Since quite a few participants believed that this was a real advertisement job they were suggesting improvements regarding the efficiency of the advertisement development. For example, a participant said: \emph{``Probably best if you removed part 3 and you guys at the sCOPa mkt department just pick an end result."} However, in general the vast majority of the participants was very happy with the job, thanking the research team. There are numerous comments on that direction: Indicatively: \emph{``Great task! one of my favorite[s] so far"} \dots \emph{``I would love to do this job again in the future! Thanks!"} \dots \emph{``Looking forward to working on more collaborative projects"}.
From the above, null hypothesis $H_{o2}$ is partially rejected, in regards to the Communication Quality, Stress levels and End Result axes.

\begin{figure}[!h]
\centering
\fbox{
\includegraphics[width=0.9\columnwidth]{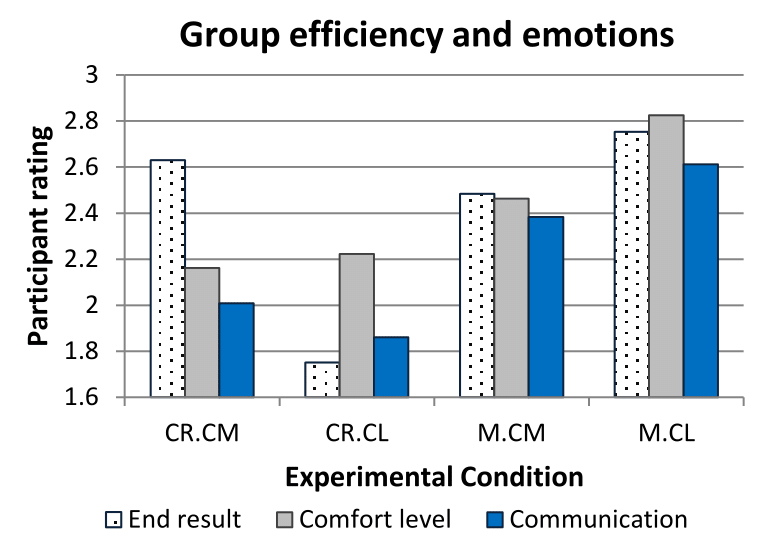}
}
\caption{Hypothesis 2 - Participant emotions during group interaction and self-perceptions of group efficiency.}
\label{fig_hyp_2}
\end{figure}
\section{Discussion, Limitations and Future Work}
\label{discussion}
Our analysis showed that the crowdsourcing population seems to have significantly higher percentages of leader (D, I) than non-leader types (S, C). Thus the probability of creating a crashing team if selecting randomly is high. This reinforces the significance of our research, enabling a more effective selection of workers through the creation of more compatible teams and thus the achievement of a higher-quality result. 

Although all participants were pleased with the cooperative nature of the tasks, people in the matching conditions and especially the collaborative ones reported better group communication, lower stress levels and liked the end result more (statistical significance of hypothesis 2). The importance of having happy and relaxed workers is sufficiently studied in organizational psychology \cite{Zeytinoglu:2005}, and it has also been indicated in crowdsourcing settings \cite{Hossain:2012}. Among the different intrinsic motivators known to affect crowdsourcing quality output (e.g. reputation, satisfaction with the task etc.), this research adds that personality matching in groups can be another powerful intrinsic motivator for work.

Groups and individuals mostly behaved as expected while interacting with the group, implying that the DISC tool has a good prognostic value. DISC also highly correlates with another well-known, valid and reliable tool, MBTI, gaining further convergence validity \cite{Schaubhut:2009}. 

The statistical significance of hypothesis 1 brings along practical benefits for crowdsourcing task designers and crowsourcing platforms. Specifically, through a relatively easy approach (a personality test and group matching) worker productivity in group tasks can be significantly increased. The same outcome can be potentially beneficial for other applications, where group tasks among previously unknown individuals can take place, such as learning applications or corporate settings.

Concerning ethics, we observed that it was very easy for people to reveal their personality traits (almost 300 responses were collected in only 2 hours). While this study strictly followed all ethical research guidelines, this is not guaranteed in commercial practices. Future work could also examine the reasons why people are ready to give their personal data and how personality can be used correctly and with integrity in crowdsourcing applications.

Finally our results are valid only for the specific task types that were studied (collaborative/competitive, creative, of short duration, relatively easy, with low leader control etc.). Other task types could be affected in different ways, or even not at all by personality matching within the group. For example, routine tasks (as opposed to creative) or tasks with high leader control, could be affected less, in the first case because personality does not need to be expressed and in the second, because the team members' roles are clearly predefined. Future work could examine the proposed approach under the scope of different task types, varying the values of the different elements presented in Table~\ref{tab:table1} of our research design.
\section{Conclusion}
\label{conclusion}
In this work we examined the impact of personality compatibility on the effectiveness of group work in cooperative crowdsourcing. Our results, on two main types of personality combinations (matching or crashing) and on two main types of tasks (collaborative and competitive), show that indeed the way people are placed together can significantly affect the final outcome of the team, as well as the emotions and satisfaction of the individual team members. Specifically we showed that teams with matching personalities perform better and are more satisfied than teams with crashing personalities. This is especially true for matching collaborative groups, although statistically significant differences were found among all four group combinations. Our results are even more important keeping in mind that in crowdsourcing, the probability of coming up with a crashing team is high, due to the high percentage of leader personalities observed in the crowd worker population. 
This work is the first to examine the effect of personality over team result in crowdsourcing settings. Its results have practical implications for crowdsourcing platforms and task designers, who want to leverage crowdsourced team work and improve its outcomes. We hope that the present research will be a first step in a new field, one that will examine personality aspects in crowdsourced group activities, and that more researchers will be inspired to continue this effort.

\balance

\bibliographystyle{acm-sigchi}
\bibliography{paper_bib}

\begin{thebibliography}{10}

\bibitem{Andre:2014}
Andr{\'e}, P., Kraut, R.~E., and Kittur, A.
\newblock Effects of simultaneous and sequential work structures on distributed
  collaborative interdependent tasks.
\newblock In {\em Proc. of the SIGCHI Conference on Human Factors in Computing
  Systems}, CHI '14, ACM (New York, NY, USA, 2014), 139--148.

\bibitem{Belbin:2010}
Belbin, R.~M.
\newblock {\em { Management Teams\: Why They Succeed or Fail }}.
\newblock Butterworth Heinemann, 3rd ed., 2010.

\bibitem{Bernstein:2011}
Bernstein, M.~S., Brandt, J., Miller, R.~C., and Karger, D.~R.
\newblock Crowds in two seconds: Enabling realtime crowd-powered interfaces.
\newblock In {\em Proc. of the 24th Annual ACM Symposium on User Interface
  Software and Technology}, UIST '11, ACM (New York, NY, USA, 2011), 33--42.

\bibitem{Brewin:1988}
Brewin, C.
\newblock {\em {Cognitive foundations of clinical psychology}}.
\newblock Lawrence Erlbaum, 1988.

\bibitem{Carver:1996}
Carver, C., and Scheier, M.
\newblock {\em {Perspectives on Personality}}.
\newblock Allyn and Bacon, 1996.

\bibitem{Cattell:1977}
Cattell, R., Eber, H., and Tatsuoka, M.
\newblock {\em {Handbook for the 16 personality factor questionnaire}}.
\newblock IPAT, 1977.

\bibitem{Chai:2009}
Chai, K., Potdar, V., and Dillon, T.
\newblock {Content Quality Assessment Related Frameworks for Social Media}.
\newblock In {\em Proc. of the International Conference on Computational
  Science and Its Applications\: Part II}, ICCSA '09, Springer-Verlag (Berlin,
  Heidelberg, 2009), 791--805.

\bibitem{Costa:1992}
Costa, P.~J., and McCrae, R.
\newblock {\em {Revised NEO Personality (NEO-PI-R) and NEO Five-Factor
  Inventory (NEO-FFI) professional manual}}.
\newblock Psychological Assessment Resources, 1992.

\bibitem{Dow:2011:PDS:1978942.1979359}
Dow, S., Fortuna, J., Schwartz, D., Altringer, B., Schwartz, D., and Klemmer,
  S.
\newblock Prototyping dynamics: Sharing multiple designs improves exploration,
  group rapport, and results.
\newblock In {\em Proc. of the SIGCHI Conference on Human Factors in Computing
  Systems}, CHI '11, ACM (New York, NY, USA, 2011), 2807--2816.

\bibitem{Downs:2010}
Downs, J.~S., Holbrook, M.~B., Sheng, S., and Cranor, L.~F.
\newblock Are your participants gaming the system?: screening mechanical turk
  workers.
\newblock

\bibitem{Eickhoff:2011}
Eickhoff, C., and de~Vries, A.~P.
\newblock How crowdsourcable is your task?
\newblock In {\em Workshop on Crowdsourcing for Search and Data Mining (CSDM)}
  (Hong Kong, China, 2011).

\bibitem{Eysenck:1967}
Eysenck, H.
\newblock {\em {The biological basis of personality.}}
\newblock Charles C Thomas, 1967.

\bibitem{Eysenck:1975}
Eysenck, H.
\newblock {\em {The inequality of man.}}
\newblock EdITS, 1975.

\bibitem{Fiedler:1983}
Fiedler, F., and Potter, E.
\newblock Dynamics of leadership effectiveness.
\newblock In {\em Small Groups and Social Interaction}, V.~K. H.H.~Blumbers,
  A.P.~Hare and M.~Davies, Eds., vol.~1, Chichester: Wiley (1983), 407--13.

\bibitem{Forlines:2014}
Forlines, C., Miller, S., Guelcher, L., and Bruzzi, R.
\newblock Crowdsourcing the future: Predictions made with a social network.
\newblock In {\em Proc. of the SIGCHI Conference on Human Factors in Computing
  Systems}, CHI '14, ACM (New York, NY, USA, 2014), 3655--3664.

\bibitem{Franzoni:2014}
Franzoni, C., and Sauermann, H.
\newblock Crowd science: The organization of scientific research in open
  collaborative projects.
\newblock {\em Research Policy 43}, 1 (2014), 1 -- 20.

\bibitem{Furnham:1992}
Furnham, A.
\newblock {\em {Personality at work}}.
\newblock Routledge, 1992.

\bibitem{Furnham:1999}
Furnham, A.
\newblock {\em {The Psychology of Behaviour at Work\: the individual in the
  organization}}.
\newblock Psychology Press, 1999.

\bibitem{Hoffman:2012}
Hoffman, B.
\newblock {The Ad Contrarian. Fowler Digital Services}.
\newblock ebook, 2012.

\bibitem{Holland:1973}
Holland, J.
\newblock {\em {Making vocational choices\: a theory of careers}}.
\newblock Prentice Hall, 1973.

\bibitem{Hossain:2012}
Hossain, M.
\newblock Users' motivation to participate in online crowdsourcing platforms.
\newblock In {\em Innovation Management and Technology Research (ICIMTR), 2012
  International Conference on} (May 2012), 310--315.

\bibitem{Josang:2007}
J{\o}sang, A., Ismail, R., and Boyd, C.
\newblock A survey of trust and reputation systems for online service
  provision.
\newblock {\em Decis. Support Syst. 43}, 2 (Mar. 2007), 618--644.

\bibitem{kargerBudget}
Karger, D.~R., Oh, S., and Shah, D.
\newblock Budget-optimal task allocation for reliable crowdsourcing systems.
\newblock {\em CoRR abs/1110.3564\/} (2011).

\bibitem{Kazai:2011}
Kazai, G., Kamps, J., and Milic-Frayling, N.
\newblock Worker types and personality traits in crowdsourcing relevance
  labels.
\newblock In {\em Proc. of the 20th ACM International Conference on Information
  and Knowledge Management}, CIKM '11, ACM (New York, NY, USA, 2011),
  1941--1944.

\bibitem{Kazai:2012}
Kazai, G., Kamps, J., and Milic-Frayling, N.
\newblock The face of quality in crowdsourcing relevance labels: Demographics,
  personality and labeling accuracy.
\newblock In {\em Proc. of the 21st ACM International Conference on Information
  and Knowledge Management}, CIKM '12, ACM (New York, NY, USA, 2012),
  2583--2586.

\bibitem{Kittur:2010}
Kittur, A.
\newblock Crowdsourcing, collaboration and creativity.
\newblock {\em XRDS 17}, 2 (Dec. 2010), 22--26.

\bibitem{Knight:2005}
Knight, S., and Burn, J.
\newblock {Developing a Framework for Assessing Information Quality on the
  World Wide Web.}
\newblock {\em Informing Science Journal 8\/} (2005), 159--172.

\bibitem{Kravitz:1986}
Kravitz, D.~A., and Martin, B.
\newblock {Ringelmann Rediscovered: The Original Article}.
\newblock {\em Journal of Personality and Social Psychology 50}, 5 (May 1986),
  936--941.

\bibitem{Lasecki:2013}
Lasecki, W.~S., Miller, C.~D., and Bigham, J.~P.
\newblock Warping time for more effective real-time crowdsourcing.
\newblock In {\em Proc. of the SIGCHI Conference on Human Factors in Computing
  Systems}, CHI '13, ACM (New York, NY, USA, 2013), 2033--2036.

\bibitem{Lord:1997}
Lord, C.
\newblock {\em {Social Psychology.}}
\newblock Harcourt Brace College Publishers, 1997.

\bibitem{Marston:1979}
Marston, W.~M.
\newblock {\em { Emotions of Normal People.}}
\newblock Persona Press Inc., 1979.

\bibitem{Miller:1956}
Miller, G.~A.
\newblock The magical number seven, plus or minus two: Some limits on our
  capacity for processing information.
\newblock {\em Psychological Review 63}, 2 (1956), 81–97.

\bibitem{Morris:2013}
Morris, R.~R., Dontcheva, M., Finkelstein, A., and Gerber, E.
\newblock Affect and creative performance on crowdsourcing platforms.
\newblock {\em 2013 Humaine Association Conference on Affective Computing and
  Intelligent Interaction 0\/} (2013), 67--72.

\bibitem{Myers:1985}
Myers, M., and McCaulley, M.
\newblock {\em {Manual\: A guide to the development and use of the Myers-Briggs
  Type Indicator}}.
\newblock Consulting Psychologists Press, 1985.

\bibitem{Nov:2013}
Nov, O., Arazy, O., L\'{o}pez, C., and Brusilovsky, P.
\newblock Exploring personality-targeted ui design in online social
  participation systems.
\newblock In {\em Proc. of the SIGCHI Conference on Human Factors in Computing
  Systems}, CHI '13, ACM (New York, NY, USA, 2013), 361--370.

\bibitem{Pekrun:2011}
Pekrun, R., Goetz, T., Frenzel, A.~C., Barchfeld, P., and Perry, R.~P.
\newblock {Measuring emotions in students’ learning and performance\: The
  Achievement Emotions Questionnaire (AEQ)}.
\newblock {\em Contemporary Educational Psychology 36}, 1 (2011), 36 -- 48.
\newblock Students\' Emotions and Academic Engagement.

\bibitem{Pervin:1984}
Pervin, L.
\newblock {\em {Current controversies and issues in personality.}}
\newblock John Wiley, 1984.

\bibitem{Ramesh:2012}
Ramesh, A., Parameswaran, A., Garcia-Molina, H., and Polyzotis, N.
\newblock Identifying reliable workers swiftly.
\newblock Technical report, 2012.

\bibitem{Sampath:2014}
Sampath, A.~H., Rajeshuni, R., and Indurkhya, B.
\newblock Cognitively inspired task design to improve user performance on
  crowdsourcing platforms.
\newblock In {\em Proc. of the 32Nd Annual ACM Conference on Human Factors in
  Computing Systems}, CHI '14, ACM (New York, NY, USA, 2014), 3665--3674.

\bibitem{Schaubhut:2009}
Schaubhut, N., Herk, N., and Thompson, R.
\newblock {\em MBTI form M\: Manual Supplement}, 2009.

\bibitem{Steiner:1972}
Steiner, I.
\newblock {\em {Group Preocesses and Productivity}}.
\newblock Academic Press, 1972.

\bibitem{Steiner:1976}
Steiner, I.
\newblock Task-performing groups.
\newblock In {\em Contemporary Topics in Social Psychology}, J.~Thibaut,
  J.~Spence, and R.~Carson, Eds., General Learning Press (1976), 393--422.

\bibitem{Taylor:1958}
Taylor, D., Berry, P., and Block, C.
\newblock Does group participation when using brainstorming facilitate or
  inhibit creative thinking.
\newblock {\em Administration Science Quarterly 3\/} (1958), 23--47.

\bibitem{Vaish:2014}
Vaish, R., Wyngarden, K., Chen, J., Cheung, B., and Bernstein, M.~S.
\newblock Twitch crowdsourcing: Crowd contributions in short bursts of time.
\newblock In {\em Proc. of the 32Nd Annual ACM Conference on Human Factors in
  Computing Systems}, CHI '14, ACM (New York, NY, USA, 2014), 3645--3654.

\bibitem{Vuurens:2011}
Vuurens, J., Vries, A. P.~D., and Eickhoff, C.
\newblock How much spam can you take? an analysis of crowdsourcing results to
  increase accuracy.
\newblock In {\em Proc. of the ACM SIGIR 2011 workshop on crowdsourcing for
  information retrieval}, CIR 2011 (2011), 48–55.

\bibitem{whitehill:2009}
Whitehill, J., Ruvolo, P., Wu, T., Bergsma, J., and Movellan, J.
\newblock {Whose Vote Should Count More: Optimal Integration of Labels from
  Labelers of Unknown Expertise}.
\newblock In {\em NIPS} (2009).

\bibitem{Wilson:1973}
Wilson, G.
\newblock {\em {The psychology of conservatism}}.
\newblock Academic Press, 1973.

\bibitem{Zeytinoglu:2005}
Zeytinoglu, I.
\newblock Satisfied workers, retained workers\: Effects of work and work
  environment on homecare workers' job satisfaction, stress, physical health,
  and retention.
\newblock Tech. Rep. RC1-0965-06, Canadian Health Services Research Foundation,
  dec 2005.

\bibitem{Zuckerman:1979}
Zuckerman, M.
\newblock {\em {Sensation-seeking\: beyond the optimal level of arousal}}.
\newblock John Wiley, 1979.

\end{thebibliography}
\end{document}